\documentclass[11pt]{article}
\usepackage{amsmath,amssymb}
\usepackage{hyperref}
\renewcommand{\Im}{\operatorname{Im}}

\newsavebox{\uuunit}
\sbox{\uuunit}
    {\setlength{\unitlength}{0.825em}
     \begin{picture}(0.6,0.7)
        \thinlines
        \put(0,0){\line(1,0){0.5}}
        \put(0.15,0){\line(0,1){0.7}}
        \put(0.35,0){\line(0,1){0.8}}
       \multiput(0.3,0.8)(-0.04,-0.02){10}{\rule{0.5pt}{0.5pt}}
     \end {picture}}

\usepackage{mathrsfs}

\newcommand{\ba}{\begin{eqnarray*}}
\newcommand{\ea}{\end{eqnarray*}}
\newcommand{\ban}{\begin{eqnarray}}
\newcommand{\ean}{\end{eqnarray}}

\newcommand{\cN}{{\cal N}}

\newcommand{\sfQ}{{\mathsf{Q}}}

\newcommand{\sfB}{{\mathsf{B}}}

\def\e{{\,\rm e}\,}

\def\dd{{\rm d}}

\def\beq{\begin{equation}}
\def\bee{\begin{equation}}
\def\eeq{\end{equation}}
\def\bea{\begin{eqnarray}}
\def\eea{\end{eqnarray}}
\def\bd{\begin{displaymath}}
\def\ed{\end{displaymath}}

\numberwithin{equation}{section}
\begin{document}

\thispagestyle{empty}
{}

%\hfill

\vskip -3mm
\begin{center}
{\bf\LARGE
\vskip - 1cm
Indefinite theta functions for counting \\ attractor backgrounds \\ [2mm]
}

\vspace{10mm}

{\large
{\bf Gabriel Lopes Cardoso$^+$,}
{\bf Michele Cirafici$^+$ and}
{\bf Suresh Nampuri$^\dagger$}

\vspace{1cm}

{\it 
$^+$
Center for Mathematical Analysis, Geometry and Dynamical Systems
\\ [1mm] 
Department of Mathematics  
\\ [1mm] 
Instituto Superior T\'ecnico, Universidade de Lisboa
\\ [1mm]
Av. Rovisco Pais, 1049-001 Lisboa, Portugal \\ [2mm]

$^\dagger$ 
National Institute for Theoretical Physics (NITheP) \\[1mm]
School of Physics and Center for Theoretical Physics \\[1mm]
University of Witwatersrand \\[1mm]
WITS 2050, Johannesburg, South Africa
}}

\vspace{5mm}

\end{center}
\vspace{5mm}

%\today

\begin{center}
{\bf ABSTRACT}\\
\end{center}

\noindent
In this note, we employ indefinite theta functions to regularize canonical partition functions
for  single-center dyonic BPS black holes.  These partition functions count dyonic degeneracies in the Hilbert space
of four-dimensional toroidally compactified  heterotic string theory, graded by electric and magnetic charges. The regularization is achieved by viewing the weighted sums of degeneracies as sums over 
charge excitations in the near-horizon attractor geometry of an arbitrarily chosen black hole background, and
eliminating the unstable modes.
This enables  us to rewrite these sums in terms of indefinite theta functions.
Background independence is then implemented by using the transformation property of indefinite theta functions
under elliptic transformations, while modular transformations are used to make contact with semi-classical results
in supergravity.

\clearpage
\setcounter{page}{1}

\tableofcontents

\section{Introduction}

Ooguri, Strominger and Vafa introduced a partition function for BPS black holes in four dimensions based on a mixed statistical ensemble \cite{Ooguri:2004zv}
\begin{equation}
{\cal Z}_{\rm OSV} (p, \phi) = \sum_{q_I} d(q, p) \, {\rm e}^{\pi q_I \phi^I} \;,
\label{OSV}
\end{equation}
where $d(q,p)$ denote microstate degeneracies that depend on electric and
magnetic charges $(q_I, p^I)$, and $\phi^I$ denote electrostatic potentials that are held fixed ($I = 0, \dots, n$).
When evaluating this partition function \cite{Shih:2005he,
LopesCardoso:2006bg,Cardoso:2008ej}, one encounters divergences that are 
associated with the indefinite signature of the
underlying lattice of electric and magnetic charges.
One therefore needs to introduce a regulator.
OSV type partition functions
have been computed in two different regimes. In one regime one utilizes the description of the system
as a bound state of D-branes and regularizes the partition function by modifying the exponent of \eqref{OSV}
through the addition of a so-called H-regulator \cite{Dabholkar:2005dt,Denef:2007vg}. 
 This calculation is done in a regime where the D-brane world sheet theory is weakly coupled.
The partition function may, however, also be computed in a different regime, 
where a supergravity description in terms of BPS black holes is available. 
In this paper we will focus on a subset of these black holes, namely single-center black holes, and 
we will be interested in single-center black hole partition functions.  To define these, the sum \eqref{OSV}
needs to be restricted in a suitable manner.  This can be done as follows.

We consider a specific model, namely four-dimensional toroidally compactified heterotic string theory.
For this model
there exists an exact counting formula of $\tfrac14$ BPS microstate degeneracies \cite{Dijkgraaf:1996it,Shih:2005uc}
in terms of a Siegel modular form $1/\Phi_{10}$, expressed as a function of quadratic charge invariants. 
To be able to use an effective $\cN=2$ description, we will work with a restricted set
of $\cN=4$ charges, which we denote by $(q_I, p^I)$.
Then, a black hole partition such as \eqref{OSV} is evaluated in various steps.  First, we sum over charges $q_0$ and $q_1$.
To do so, we express these charges in terms of T-duality invariant charge bilinears, and we rewrite 
the chemical potentials $\phi^0$ and $\phi^1$ in \eqref{OSV} in terms of the Siegel upper half plane period matrix entries, which act as chemical potentials for the charge invariants that  parametrize the degeneracies in the ensemble counted by $1/\Phi_{10}$. This allows us to express the sum over $q_0$ and $q_1$ in terms of 
an integral over $1/\Phi_{10}$, which is then evaluated using residue techniques.
In doing so we restrict
the analysis to a certain subset of zeroes of $\Phi_{10}$ \cite{David:2006yn}, in order to 
single out the contributions
that give rise to the dilatonic free energy ${\cal F}_D$ of a single-center black hole \cite{LopesCardoso:2006bg}.
Subsequently, we also sum over charges $q_a$ and $p^a$, obtaining generalized OSV type partition functions which
we call single-center black hole partition functions. 
For extremal dyonic black holes, the near-horizon geometry, called the attractor geometry (for reasons explicated in the next section), decouples from asymptotic infinity and 
encodes the entropy of the black hole microstates sans scalar hair contributions. The single-center black hole  partition functions that we obtain\footnote{These partition functions are different from the finite part of the index, $\psi_m^F (\tau, z)$, defined in \cite{Dabholkar:2012nd}, that counts states in the CFT dual to the near-horizon geometry, in an ensemble parametrized by charge invariants at fixed magnetic charges.} count excitations,
graded by $q_a$ and $p^a$, in the near-horizon geometry of the black hole.

When performing the sum over charges $q_a$ (and $p^a$) we encounter the
aforementioned divergences associated with the indefiniteness of the $q_a (p^a)$ charge lattice.
In   
\cite{Cardoso:2013ysa} we advocated using indefinite theta functions \cite{Zwegers:2008} to 
regularize these sums.
There we focussed on OSV black hole partition functions with $p^0=0$.
In this paper we will extend our considerations and consider single-center black hole partition functions 
with $p^0 \neq 0$ based on either mixed
or canonical ensembles.  To regularize the sum over $q_a$ (and $p^a$) we first pick a reference attractor
background (which we define
in next the section) and consider fluctuations in this background.  To enforce thermodynamical stability, we restrict
to fluctuations that do not increase the dilatonic free energy ${\cal F}_D$ mentioned above. Thus, we 
remove exponentially growing contributions.  This is done by introducing in the sum
a suitable measure factor based on sign functions (rather than by modifying the exponent of \eqref{OSV}).
The resulting
regularized sums are given in terms of indefinite theta functions.  The latter have 
good transformation properties under modular and elliptic transformations.  The elliptic transformation
property ensures that the result is independent of the chosen reference background. 
We use modular transformations to extract known semi-classical results from the
regularized partition functions, namely the semi-classical free energy ${\cal F}_E$ and the 
semi-classical Hesse potential ${\cal H}$ \cite{Ooguri:2004zv,LopesCardoso:2006bg}.
We note that this regularization procedure requires, in addition, extending the electrostatic potentials $\phi^a$ to complex
potentials $\phi^a + i \mu^a$, and similarly for their magnetic counterparts $\chi_a$ ($a= 2, \dots, n$),
as was already noted in \cite{Cardoso:2013ysa}. Indefinite theta functions have previously found applications
in counting dyonic degeneracies \cite{Manschot:2009ia,Manschot:2010xp,Alexandrov:2012au}.

This paper is organized as follows. In section \ref{bkgd} we introduce the notion of an attractor background
and collect various useful formulae.  In section \ref{bh-sec-pf} we define single-center black hole partition functions
for two types of ensembles
in toroidally compactified heterotic string theory, and we describe the regulator that we use to deal with the
aforementioned divergences.  Contrary to \cite{Cardoso:2013ysa} we do not restrict to single-center
black holes with $p^0 =0$.  The resulting regularized partition functions have 
good modular and elliptic transformation properties, which we use to make contact with semi-classical supergravity
results.  We summarize our findings in section \ref{discussion}, where we also comment on various subtleties
that we encountered.

\section{Attractor backgrounds}\label{bkgd}

We consider models whose two-derivative Wilsonian effective action is based on an 
$\cN=2$ prepotential of the form
\begin{equation} \label{eq:heterotic-prep}
F^{(0)} (Y) = - \frac12 \frac{Y^1 Y^a C_{ab} Y^b}{Y^0} \qquad , \qquad a=2,\dots,n \;,
\end{equation}
up to worldsheet instanton corrections which we assume are either absent or negligible. 
Here, $n$ denotes the number of $\cN=2$ abelian vector multiplets coupled to $\cN=2$ supergravity.
The consistent coupling of these vector multiplets to supergravity
requires the symmetric matrix $C_{ab}$ to have signature $(1,n-2)$ \cite{Ferrara:1989py,Aspinwall:2000fd}.  These models may either describe
genuine $\cN=2$ models or provide an effective $\cN=2$ description of $\cN=4$ models when restricting to
a subset of $\cN=4$ charges.

We introduce the quantity $K^{(0)} = i \left( \bar{Y}^I F^{(0)}_I - Y^I {\bar F}_I^{(0)} \right)$, where $I = 0, 1, \dots, n$ and $F_I^{(0)}  = \partial F^{(0)} (Y)/\partial Y^I $. For the class of models specified by \eqref{eq:heterotic-prep},  $K^{(0)}$ takes the form
\begin{equation}
K^{(0)} = \frac12 |Y^0|^2 (S + \bar S) (T+ \bar T)^a C_{ab} (T + \bar T)^b \;,
\label{K-value}
\end{equation}
where we defined
\begin{equation}
S = -i \frac{Y^1}{Y^0} \;\;\;,\;\;\; T^a = -i \frac{Y^a}{Y^0}\;.
\end{equation}
In heterotic string theory, the field $S$ denotes the dilaton/axion complex scalar field.

We can construct 
single-center dyonic BPS black hole solutions in any given model \eqref{eq:heterotic-prep}.  These are static, spherically symmetric, asymptotically Minkowskian spacetimes with line elements given by
\begin{equation}
\dd s^2 = - e^{2 U(r)} \dd t^2 + e^{- 2 U(r)} \left( \dd r^2 + r^2  \dd \Omega^2_{(2)} \right)\ .
\end{equation}
These solutions, which are supported by scalar fields $Y^I(r)$ and by the abelian gauge fields of the model,
are dyonic and carry electric/magnetic charges $(q_I , p^I)$.

A fixed charge vector $(q_I , p^I)$ supports a single-center
BPS black hole solution if the scalar fields $Y^I(r)$ evolve smoothly to near-horizon values 
$Y^I = Y^I_*/r$ specified by the 
so-called attractor equations \cite{Ferrara:1995ih,Ferrara:1996dd,Ferrara:1996um,Behrndt:1996jn}
\begin{eqnarray}
Y^I_* - \bar{Y}^I_* &=& i p^I \ , \nonumber\\
F_I^{(0)}(Y_*) - \bar{F}_I^{(0)}({\bar Y}_*) &=& i q_I \;,
\label{attract-eqs}
\end{eqnarray}
such that the horizon quantity $|Z_*|^2 \equiv  p^I F^{(0)}_I (Y_*) - q_I Y^I_*$ is non-vanishing, i.e.
$|Z_*|^2 >0$.  Then,
the near-horizon line element takes the form of an $AdS_2 \times S^2$ line element, 
\begin{equation}
\dd s^2  = - \frac{r^2}{|Z_{*}|^2} \dd t^2 + \frac{|Z_{*}|^2}{r^2} \, \dd r^2 + |Z_{*}|^2 \, \dd \Omega^2_{(2)} \ ,
\end{equation}
and the macroscopic entropy of the BPS black hole, which at the two-derivative level is determined by
the area law, equals
${\cal S}(q,p) = \pi \, |Z_{*}|^2 = \pi \left(  p^I F^{(0)}_I (Y_*) - q_I Y^I_* \right)$.
The entropy may also be expressed as ${\cal S}(q,p) = \pi K^{(0)}$ by virtue of \eqref{attract-eqs}.

Next, we associated a free energy to the BPS black hole.  To this end, we introduce 
electro/magnetostic potentials as \cite{LopesCardoso:2000qm}
\begin{eqnarray}
\phi^I &=& Y^I + \bar{Y}^I \ ,
\nonumber\\
\chi_I &=& F^{(0)}_I + \bar{F}^{(0)}_I \ .
\end{eqnarray}
Then, a solution to the attractor equations \eqref{attract-eqs} can be expressed as
\begin{eqnarray}
Y^I_* &=& \frac12 \left( \phi^I_* + i p^I  \right) \ , 
\label{Y-p}\\
F^{(0)}_I (Y_*) &=& \frac12 \left( \chi_{I*} + i q_I \right) \ .
\label{F-q}
\end{eqnarray}
The black hole can be assigned a macroscopic free energy by performing a Legendre transform of the entropy.
There are various possibilities here.
Performing a Legendre transform with respect to all the electric charges yields the free energy
${\cal F}_E^{(0)}(p, \phi_*) = {\cal S}(q,p)/\pi + q_I \phi^I_*$, which equals \cite{Ooguri:2004zv,LopesCardoso:2006bg}
\begin{eqnarray}
{\cal F}_E^{(0)}(p, \phi_*) &=& 4 \left[ {\rm Im} F^{(0)}(Y) \right]\Big{\vert}_{Y^I_* = \tfrac12 (\phi^I_* + i p^I)}
\nonumber\\
&=& \frac14 (S + \bar S) \left[ p^a C_{ab} p^b - \phi^a_* C_{ab} \phi^b_* - 2 i \frac{S- \bar S}{S + \bar S} \,
\phi^a_* C_{ab} p^b \right] \;,
\label{electric-free}
\end{eqnarray}
where $S$ is expressed in terms of
the electrostatic potentials $\phi^0_*$ and $\phi^1_*$ and the magnetic charges $p^0,p^1$ as
\begin{equation}
S = \frac{- i \phi^1_* + p^1}{\phi^0_* + i p^0} \;.
\end{equation}
On the other hand, performing the Legendre transform with respect to the electric charges $q_0, q_1$ only
yields the dilatonic free energy ${\cal F}_D^{(0)}(S, \bar S, p^a, q_a) = 
{\cal F}_E^{(0)}(p, \phi_*) - q_a \phi^a_*$,
which equals \cite{LopesCardoso:2006bg}
\begin{eqnarray}
{\cal F}^{(0)}_D (S, \bar S, p^a, q_a) = 
\frac{1}{S + \bar S} \left[ q_a C^{ab} q_b + |S|^2 p^a C_{ab} p^b +  i (S- \bar S) q_a p^a
\right].
\label{FD0}
\end{eqnarray}
Finally, performing a Legendre transform of  ${\cal F}_E^{(0)}(p, \phi_*) $
with respect to the magnetic charges $p^a$
yields the reduced Hesse potential ${\cal H}^{(0)}(S, \bar S, \phi^a_*, \chi_{a*})=
{\cal F}_E^{(0)}(p, \phi_*) - p^a \chi_{a*}$
\cite{Cardoso:2013ysa},
\begin{equation}
{\cal H}^{(0)}(S, \bar S, \phi^a_*, \chi_{a*}) = - \frac{1}{S + \bar S} \left[
\chi_{a*} C^{ab} \chi_{b*} + |S|^2 \phi^a_* C_{ab} \phi^b_* + i (S - \bar S) \chi_{a*} \phi^a_*  \right] \;.
\label{hesse-a}
\end{equation}
The extremization equations following from 
 ${\cal H}^{(0)}(S, \bar S, \phi^a_*, \chi_{a*})=
 {\cal F}^{(0)}_D (S, \bar S, p^a, q_a) + q_a \phi^a_* - p^a \chi_{a*}$
 yield the 
attractor values $(\phi^a_*, \chi_{a*})$,  expressed 
in terms of 
the charges $(q_a, p^a)$ and the field $S$, namely
\begin{eqnarray}
&&\phi^a_* + 2\, \frac{C^{ab} \, q_b}{S + {\bar S}} +  \frac{i(S - \bar S)}{S + \bar S} \, p^a= 0 
\label{eq:phia-attrac}
\end{eqnarray}
and
\begin{equation}
\chi_{a*} -2 \frac{|S|^2}{S + \bar S}  C_{ab} p^b
-
i \frac{(S - \bar S)}{S + \bar S} q_a  = 0 \;.
\label{eq:chia-attrac}
\end{equation}
At the two-derivative level, the attractor value of $S$ is determined in terms of the following three charge bilinears
\begin{equation}
Q = 2 q_0 p^1 - q_a C^{ab} q_b \;\;\;,\;\;\;
P = - 2 p^0 q_1 - p^a C_{ab} p^b \;\;\;,\;\;\; R = p^0 q_0 - p^1 q_1 + p^a q_a 
\label{charge-bilinears}
\end{equation}
as \cite{LopesCardoso:1999ur}
\begin{equation}
S = \sqrt{\frac{Q P - R^2}{P^2}} -i \frac{R}{P} \;.
\end{equation}
The entropy, when expressed in terms of these charge bilinears, reads ${\cal S}(q,p) = 
\pi \, \sqrt{Q P - R^2}$.

Conversely, given a value $S$ with $S + \bar S > 0$ and charges $(q_a, p^a)$, we defined attractor values $(\phi^a_*, \chi_{a*})$
by \eqref{eq:phia-attrac} and \eqref{eq:chia-attrac}.  Therefore, the lattice of electric and magnetic
charges singles out a subset of values $(\phi^a, \chi_a)$, namely the attractor values $(\phi^a_*, \chi_{a*})$.
For a given $S$, these correspond to attractor values
\begin{equation}
Y^a_* =  \frac{\bar S \, \phi^a_*  + i C^{ab} \chi_{b*}}{S + \bar S} \;.
\label{Y-attrac}
\end{equation}
In the following, 
we will refer to the attractor values $(\phi^a_*, \chi_{a*})$ as attractor backgrounds, provided $Q P - R^2 >0$
as well as $\varrho_a \, C^{ab} \varrho_b >0$. The latter are necessary conditions for a charge
configuration to constitute a single-center BPS black hole, as we show below.  
Before doing so, we note that
all three quantities ${\cal F}_E^{(0)}(p, \phi_*), {\cal F}^{(0)}_D (S, \bar S, p^a, q_a),
{\cal H}^{(0)}(S, \bar S, \phi^a_*, \chi_{a*})$ will play a role at various steps when 
evaluating OSV type partition functions in the following sections.

Next, let us introduce the vector $\varrho_a$, which can be motivated as follows.
Let us return to $K^{(0)}$ given in \eqref{K-value}.  Imposing the magnetic attractor equations \eqref{Y-p}
as well as the electric attractor equations \eqref{eq:phia-attrac} for the $q_a$ 
results in 
\begin{equation}
Y^0_* = \frac{p^1 + i {\bar S} p^0}{S + \bar S} \;\;\;,\;\;\; T^a_* = i \,\frac{(C^{ab} q_b - i \bar S \, p^a)}{p^1 + i \bar S p^0} \;,
\label{Y0T}
\end{equation}
and 
determines $K^{(0)}$ in terms of
the charges $(q_a, p^I)$ and $S$ as 
\begin{equation}
K^{(0)}(S, \bar S, p^I, q_a) =\frac{\varrho_a C^{ab} \varrho_b }{
2 |Y^0_*|^2 (S + \bar S)}\;, 
\label{K0-rho}
\end{equation}
where
\begin{equation}
\varrho_a = p^0 \, q_a + p^1 \, C_{ab} \, p^b \;.
\label{rho-charges}
\end{equation}
Now let us recall that the entropy of a single-center BPS black hole is given by
${\cal S}(q,p) = \pi \, K^{(0)}$, which implies 
$\varrho_a \, C^{ab} \varrho_b > 0$ in order for the entropy to be non-vanishing at the two-derivative level 
(here we are assuming $S + \bar S > 0$, with $|Y^0_*|^2 (S + \bar S)$ finite). 
Thus, $\varrho_a \, C^{ab} \varrho_b > 0$ is a necessary condition for a charge configuration to correspond
to a single-center black hole.
This combination may be expressed in terms of the charge bilinears \eqref{charge-bilinears} as
\begin{equation}
\varrho_a \, C^{ab} \varrho_b = - (p^0)^2 \, Q - (p^1)^2 \, P + 2 p^0 p^1 \, R \;.
\label{rhoQPR}
\end{equation}
Single-center black hole solutions necessarily have $QP - R^2 >0$.  We will now show, using \eqref{rhoQPR},
that they also
have to satisfy
$Q< 0 , P< 0$.  This can be checked as follows. Since  $QP - R^2 >0$, we only have two possibilities: either
$Q<0, P<0$ or $Q>0, P>0$.
Let us first assume that $Q<0, P<0$, in which case we may 
rewrite \eqref{rhoQPR} as
\begin{equation}
\varrho_a \, C^{ab} \varrho_b = \left( p^0 \, \sqrt{|Q|} \pm p^1 \, \sqrt{|P|} \right)^2
+ 2 p^0 p^1 \left(R \mp\sqrt{|QP|} \right) \;.
\end{equation}
When $p^0 p^1 > 0$, we choose the plus sign in the second term, while when $p^0p^1 <0$ we take the minus sign.
Then, using $\sqrt{|QP|} > |R|$, we obtain $\varrho_a \, C^{ab} \varrho_b >0$.
Now let us consider the case when
$Q >0, P>0$.  We rewrite 
\eqref{rhoQPR} as
\begin{equation}
\varrho_a \, C^{ab} \varrho_b = - \left( p^0 \, \sqrt{Q} \pm p^1 \, \sqrt{P}\right)^2
+ 2 p^0 p^1 \left(R \pm\sqrt{QP} \right) \;.
\end{equation}
When $p^0p^1 >0$, we choose the minus sign in the second term, while we choose the plus sign when $p^o p^1 <0$.
Using $\sqrt{|QP|} > |R|$, we see that $\varrho_a \, C^{ab} \varrho_b <0$, which establishes that configurations
with $Q >0, P>0$ cannot correspond to single-center black holes.

In section \ref{bh-sec-pf}, we will find it useful to perform the replacements
\begin{equation}
q_a \rightarrow - \frac{i}{\pi} \, \frac{\partial}{\partial \mu^a} \;\;\;,\;\;\; p^a \rightarrow \frac{i}{\pi} \frac{\partial}{\partial \nu_a} \;,
\label{diff-qp}
\end{equation}
in \eqref{Y0T},
resulting in differential operators
\begin{eqnarray}
\hat{T}^a = \pi^{-1} \frac{ C^{ab} \partial/\partial \mu^b + i {\bar S} \partial/\partial \nu^a }{(S + \bar S) Y^0_*} 
\label{hat-T}
\end{eqnarray}
and
\begin{equation}
\hat{K}^{(0)} = \frac12 |Y^0_*|^2 (S + \bar S) (\hat{T}+ \bar{\hat{T}})^a C_{ab} (\hat{T} + \bar{\hat{T}})^b \;.
\end{equation}

Observe that $K^{(0)}$ and $\hat{K}^{(0)}$ are  invariant under S-duality, provided the differential operators in 
\eqref{diff-qp} transform in the same way as the charges $(q_a, p^a)$.  Under S-duality, $S$ transforms as
\begin{equation}
S \rightarrow \frac{ a S - i b }{i c S + d} \;,
\label{Sduality}
\end{equation}
with $a, b, c, d \in \mathbb{Z}$ satisfying $ad-bc =1$, 
while the charges transform as
\begin{equation}
        \begin{array}{rcl}
      p^0 &\to& d \, p^0 + c\, p^1 \;,\\
      p^1 &\to& a \, p^1 + b \, p^0 \;,\\
      p^a &\to& d\, p^a - c \,C^{ab} \, q_b  \;,
    \end{array}
    \quad
    \begin{array}{rcl}
      q_0 &\to& a\,  q_0 -b\,q_1 \;, \\
      q_1 &\to& d \,q_1 -c\, q_0 \;,\\
      q_a &\to& a \, q_a -b\,  C_{ab} \, p^b \;.
    \end{array}
    \label{eq:electro-magn-dual_charges}
\end{equation}
The electric and magnetic potentials $\phi^I$ and $\chi_I$ transform in a similar manner, and hence the combination $q_I \, \phi^I - p^I \, \chi_I$ is invariant under S-duality.
It also follows that  $Y^0_* \rightarrow (d + i c S) \, Y^0_*$, and that $|Y^0_*|^2 (S + \bar S)$ and $T^a_*$ are invariant under S-duality.

We will also introduce the quantities $K$ \cite{LopesCardoso:2006bg},
\begin{eqnarray}
K = \frac12 |Y^0|^2 (S + \bar S) \Big[ ( T + \bar{T})^a C_{ab} ( T + \bar{T})^b
+ 4 \frac{\partial_S \Omega}{(Y^0)^2} + 4 \frac{\partial_{\bar S} \Omega}{(\bar Y^0)^2} \Big] \;,
\label{K-Om}
\end{eqnarray}
where $\Omega$ denotes a real quantity that encodes corrections due to higher-curvature terms. 
Aspects of the sigma-model geometry based on \eqref{K-Om} have been discussed in \cite{Cardoso:2012mc}.
In the context of the 
$\cN=4$ model which we will
be considering, $\Omega$ only depends on $S$ and $\bar S$ and is S-duality invariant. Hence, $K$ is S-duality invariant.
It is also T-duality invariant \cite{LopesCardoso:2006bg}.   Replacing $T^a$ by the differential operator \eqref{hat-T} yields $\hat{K}$,
which will play the role of a measure factor in subsequent discussions.

\section{Single-center black hole partition functions \label{bh-sec-pf}}

In this section we focus on a particular $\cN=4$ model for which there exists an exact counting formula for $\frac14$
BPS microstates, namely four-dimensional toroidally compactified heterotic string theory \cite{Dijkgraaf:1996it,Shih:2005uc}.
We restrict to a subset of $\cN=4$ charges, which we denote by $(q_I, p^I)$ (with $I=0, 1, \dots, n$), so as to
use an effective $\cN =2$ description of this model based on a prepotential of the form
\eqref{eq:heterotic-prep}.  The charges $(q_I, p^I)$ and the matrices $C_{ab}$ and $C^{ab}$ are integer valued, and thus
the charge bilinears \eqref{charge-bilinears} satisfy $Q, P  \in 2 \mathbb{Z}, \, R \in \mathbb{Z}$.
The BPS microstate degeneracies $d(q,p)$ are encoded in a Siegel modular form, defined on the Siegel upper half-plane $(\sigma, \rho, v)$ with  ${\rm Im} \, \sigma > 0, \; {\rm Im} \, \rho > 0, \;  ({\rm Im} \, \sigma)
({\rm Im} \, \rho ) > ({\rm Im} \, v)^2$, 
\begin{equation}
\frac{1}{\Phi_{10} (\sigma, \rho, v)} = \sum_{Q, P \leq 2, \, R \in \mathbb{Z}}
d(Q, P, R) \, {e}^{- \pi i (Q \, \sigma + P \, \rho + R \, (2v -1))} \, 
\label{siegel-rep}
\end{equation}
Convergence of the $Q$ and $P$ sums is enforced by ${\rm Im} \, \sigma >0 \,, {\rm Im} \, \rho >0$. 
The sum over $R$ is more subtle.  Convergence of the $R$ sum requires restricting it to
a certain range, and this range depends 
on the sign of ${\rm Im} \, v $
\cite{Jatkar:2005bh,Dabholkar:2007vk}.

In the following, we will focus on single-center $\frac14$ BPS black holes with $p^0 \neq 0$ and
define an OSV black hole partition function \eqref{OSV} for these in a two-step procedure, as follows.
The first step is implemented by considering the sum over electric charges $q_0, q_1$,
\begin{equation}
\sum_{q_0, q_1} d(q, p) \, {\rm e}^{\pi q_I \phi^I} \;,
\label{q0q1-sum}
\end{equation}
converting it into a sum over $Q$ and $P$ by using the relations \eqref{charge-bilinears}, 
and subsequently using an
 integral representation for the degeneracies $d(q,p)$ based on \eqref{siegel-rep}, which is
then computed in terms of residues associated with the zeros of $\Phi_{10}$.
Here we improve on the analysis of \cite{Cardoso:2013ysa} by 
only retaining those zeroes of $\Phi_{10}$ that give a contribution
to the dilatonic free energy \eqref{FD0}.

In a second step we sum over charges $q_a$. 
We begin by picking reference charges $q_a^B$ (and $p^a$) which we encode in 
a reference vector \eqref{rho-charges},
denoted by $\varrho_a^B$, 
that satisfies $\varrho_a^B C^{ab} \varrho_b^B > 0$. This is a necessary condition for the configuration
to correspond to a single-center black hole, as already discussed. 
We will refer to $\varrho_a^B$ as an attractor background.
We then consider fluctuations $V_a = q_a - q_a^B$  around
this background.
The sum over these electric fluctuations is ill defined due to the indefiniteness of the
charge lattice.  We regularize this sum by removing all the contributions  
that grow
exponentially. In addition, we demand that the resulting regularized sum possesses good transformation properties
under modular and elliptic transformations.
One way of achieving this is to convert the sum over $V_a$ into
an indefinite theta function
\cite{Cardoso:2013ysa}.  
Indefinite theta functions have good modular and elliptic transformations properties 
\cite{Zwegers:2008} which we subsequently utilize to make contact with semi-classical results. 
The regularized partition function then contains an exponential factor that accounts for
the semi-classical free energy of the background, as well as an indefinite theta functions
that describes a regularized sum of fluctuations around the background.
Due to the elliptic property of the indefinite theta function, the result is actually independent
of the choice of the background charge $q_a^B$, since two such choices are related by an elliptic transformation.
The result for the partition function also uses a differential operator that enforces
the condition $\varrho_a C^{ab} \varrho_b >0$ discussed below \eqref{rho-charges}.  Thus, 
the partition function can be viewed as a sum over attractor backgrounds.

Subsequently,
we extend the discussion by considering single-center black hole partition functions based on a canonical ensemble,
obtained by also 
summing over magnetic charges $p^a$,
\begin{equation}
{\cal Z} (p^0, p^1, \phi^I, \chi_a) = \sum_{q_I, p^a} d(q, p) \, {\rm e}^{\pi [ q_I \phi^I - p^a \chi_a] } \;.
\label{Z-enlarg-ens}
\end{equation}
We restrict our analysis to the case $S = \bar S$, so as to 
decouple the sums over $(q_a, p^a)$. We 
regularize these sums by again employing indefinite theta functions. After resorting to modular transformations,
the resulting expression is given in terms of the Hesse potential \eqref{hesse-a}, two indefinite
theta functions and a measure factor, and it 
is invariant under the strong-weak coupling duality transformation $S \rightarrow 1/S$.

We proceed to explain these results.

\subsection{Summing over charges $q_I$}

We begin by considering the sum over charges $(q_0, q_1)$, using various results obtained in \cite{Cardoso:2013ysa}.
In doing so, we improve on the analysis of \cite{Cardoso:2013ysa} and 
clarify certain statements made there. There, we specialized to $p^0=0$.  Here,
we keep $p^0 \neq 0$ (as well as $p^1 \neq 0$), which will be kept fixed throughout.

As stated above, we focus on toroidally compactified heterotic string theory, for which there exists an exact counting
formula for $\frac14$BPS microstates based on the Siegel modular form $1/\Phi_{10}$.  
We first convert the sum over $(q_0, q_1)$
into a sum over the charge bilinears $(Q, P)$ 
using the relations \eqref{charge-bilinears}, where we keep $(q_a, p^a)$ fixed.
{From} \eqref{siegel-rep} we see that the states that contribute are states for which $Q$ and $P$ are mostly negative, which implies that the states contributing to the sum over $(q_0, q_1)$ are mostly states with a definite sign
of $(q_0, q_1)$. Replacing $(q_0, q_1)$ by  $(Q, P)$ we obtain $q_0 \phi^0 + q_1 \phi^1 = 
\frac12 [Q \, \phi^0 p^1/(p^1)^2 - P \, \phi^1 p^0/(p^0)^2 ] + \dots$, where the dots refer to terms that
do not involve $Q$ and $P$. Thus,
for the exponent in \eqref{q0q1-sum} to be damped for negative $Q$ and $P$, we require 
\begin{eqnarray}
\phi^0 p^1 > 0 \;\;\;,\;\;\; \phi^1 p^0 < 0 \;.
\label{phi0phi1-cond}
\end{eqnarray}
Introducing
\begin{equation}
S = \frac{- i \phi^1 + p^1}{\phi^0 + i p^0} \;,
\label{S-phi0phi1}
\end{equation}
and using \eqref{phi0phi1-cond}, we obtain 
\begin{equation}
S + \bar S =2  \frac{(\phi^0 p^1 - \phi^1 p^0)}{|\phi^0 + i p^0|^2} > 0 \;.
\end{equation}

Next, let us consider the combination $R$ in \eqref{charge-bilinears}.
Replacing $(q_0, q_1)$ by $(Q, P)$ we obtain the combination
\begin{equation}
R(Q, P) = \frac{p^0}{2 p^1} \left(Q + q_a C^{ab} q_b \right) + \frac{p^1}{2 p^0} \left(P + p^a C_{ab} p^b \right)
+ q_a p^a \;.
\end{equation}
For fixed $(q_a, p^I)$, and taking $|Q| \gg 1, \, |P| \gg1$, the sign of $R(Q,P)$ equals the sign of $- p^0 p^1$.
Next, we convert the sum \eqref{q0q1-sum} over $(q_0, q_1)$ into a sum over $(Q,P)$ following
\cite{Shih:2005he,LopesCardoso:2006bg}.  In order to use the representation
\eqref{siegel-rep}, we introduce an additional sum over a dummy variable $R' \in \mathbb{Z}$ \cite{Cardoso:2013ysa},
\begin{equation}
f(R) = 
 \int_0^1 d \theta_1 \, \sum_{R'} \,
 e^{\pi i (2 \theta -1)  (R - R')} \, f(R') \, ,
\end{equation}
where $\theta = \theta_1 + i \theta_2 \in \mathbb{C}$, and where $\theta_2$ is held fixed with $\theta_2 \neq 0$.  For a fixed $\theta_2$, convergence of the $R'$ sum requires restricting
it to a certain range that is taken to include $R$.  This is similar to what was observed
below \eqref{siegel-rep}. 

Using this, we 
obtain the following 
representation for the sum over $(q_0, q_1)$ \cite{Cardoso:2013ysa},
\begin{eqnarray}
\sum_{q_0, q_1} d(q,p) \, {\rm e}^{\pi q_I \phi^I   } 
&=& \frac{1}{|p^0 p^1|}  \sum_{\tiny \begin{array}{c} l^0= 0 , \dots |p^1| -1\\ 
 l^1= 0 , \dots |p^0| -1 \end{array} }\,   \int_0^1 \dd \theta_1 
\, \frac{1}{\Phi_{10} (\sigma(\theta), \rho (\theta) , v(\theta))} \nonumber\\
&& \qquad {\rm exp} \left[ - i \pi \sigma(\theta)\, q_a C^{ab} q_b + \pi q_a {\tilde \phi}^a (\theta) - \pi i \, \rho (\theta) \, p^a C_{ab} p^b 
\right] \,,
\label{eq:Z-phi-theta}
\end{eqnarray}
where 
\begin{eqnarray}
\sigma (\theta) &=& i \, \frac{\hat{\phi}^0}{2 p^1} - (2 \theta -1) \, \frac{p^0}{2p^1} \;, \nonumber\\
\rho (\theta) &=& - i \, \frac{\hat{\phi}^1}{2 p^0} - (2 \theta -1) \, \frac{p^1}{2p^0} \;, \nonumber\\
v (\theta) &=& \theta \;, 
\label{eq:srvthet}
\end{eqnarray}
and 
\begin{eqnarray}
\hat{\phi}^0 &=& \phi^0 + 2 i l^0 \;,\nonumber\\
\hat{\phi}^1 &=& \phi^1 + 2 i l^1 \;,\nonumber\\
{\tilde \phi}^a (\theta) &=& \phi^a + i \, (2 \theta -1) \, p^a \;.
\end{eqnarray}
The extra sum over the integers $l^0$ and $l^1$ arises when trading the summation variables $(q_0 , q_1)$ for the T-duality invariant combinations $Q$ and $P$ \cite{Shih:2005he,LopesCardoso:2006bg}. 
The integration contour in \eqref{eq:Z-phi-theta} is at fixed $\theta_2$, whose 
value 
is  obtained by requiring that 
the conditions for convergence of the expansion \eqref{siegel-rep} in the Siegel upper half plane
are satisfied when restricting $\sigma$ and $\rho$ to \eqref{eq:srvthet}.  Namely, demanding 
${\rm Im} \, \sigma (\theta) >0, {\rm Im} \, \rho (\theta) >0$, we obtain
\begin{eqnarray}
\label{eq:phicond}
p^0 p^1 \theta_2 &<& \frac{\phi^0 p^1}{2} \;, \nonumber\\
p^0 p^1 \theta_2 &< & - \frac{\phi^1 p^0}{2}  \;, 
\end{eqnarray}
from which we infer
\begin{eqnarray}
p^0 p^1 \theta_2 < \frac{(S + \bar{S}) | \phi^0 + i p^0|^2}{8}  \;,
\label{o2-s-p}
\end{eqnarray}
where $S$ is given in \eqref{S-phi0phi1}.
Recalling \eqref{phi0phi1-cond}, we see that the right hand side of
\eqref{eq:phicond} is positive.  Taking it to be very large, so that 
$(S + \bar{S}) | \phi^0 + i p^0|^2$ is very large, we see
that 
the conditions \eqref{eq:phicond} (as well as \eqref{o2-s-p}) 
are satisfied for any finite value of $\theta_2$.
Similar considerations apply to the Siegel upper half plane condition $\Im \sigma (\theta) \;
\Im \rho (\theta) >  (\Im v(\theta))^2$,
which translates into 
\begin{eqnarray} \label{Supperhalf}
p^0 p^1 \theta_2 < \frac{-\phi^0 \phi^1 p^0 p^1}{
(S + \bar{S}) | \phi^0 + i p^0|^2}
 \;.
\end{eqnarray}

Now we note that we can also impose the more restrictive condition ${\rm Im} \, \sigma (\theta) \gg 1, {\rm Im} \, \rho (\theta) \gg 1$, which ensures that (\ref{siegel-rep}) has 
a well defined expansion for very large charges. We obtain
\begin{eqnarray}
\label{eq:phicond-sc}
p^0 p^1 \theta_2 &\ll& \frac{\phi^0 p^1}{2} - p_1^2 \;, \nonumber\\
p^0 p^1 \theta_2 &\ll & - \frac{\phi^1 p^0}{2} - p^2_0 \;, 
\end{eqnarray}
from which it follows that
\begin{eqnarray}
p^0 p^1 \theta_2 < \frac{(S + \bar{S}) | \phi^0 + i p^0|^2}{8} - \frac12 \left(p^2_0 + p^2_1 \right) \;,
\label{o2-s-p-sc}
\end{eqnarray}
In this case the conditions 
 \eqref{eq:phicond-sc} and \eqref{o2-s-p-sc} 
can be satisfied for any finite value of $\phi^0, \phi^1, p^0, p^1$
by taking $\theta_2$ to satisfy $p^0 p^1 \theta_2 <0$ with $|\theta_2| \gg 1$.  This choice also ensures
the validity of the Siegel upper half plane condition (\ref{Supperhalf}). Thus 
\begin{equation}
p^0 p^1 \theta_2 <0 \ \text{ with } \ |\theta_2| \gg 1
\label{attractorcontour}
\end{equation}
specifies another viable integration contour for the integral \eqref{siegel-rep}. Below we will show that
this choice of contour is necessary in order to select large charge single-center black holes.

The left hand side of \eqref{eq:Z-phi-theta} is invariant under the shifts $\phi^0 \rightarrow \phi^0 + 2 i \;,\; 
\phi^1 \rightarrow \phi^1 + 2i$.  The right hand side of  \eqref{eq:Z-phi-theta} is also invariant under these shifts.
This follows from the fact that the integrand of \eqref{eq:Z-phi-theta} is invariant under shifts
\begin{eqnarray}
\sigma (\theta) \rightarrow \sigma (\theta) - n \;\;\;,\;\;\; \rho (\theta) \rightarrow \rho (\theta) + m \;\;\;,\;\;
n, m \in \mathbb{Z} \;,
\label{sr-shift}
\end{eqnarray}
which are induced by 
\begin{eqnarray}
\phi^0 \rightarrow \phi^0 + 2i p^1 \, n \;\;\;,\;\;\; \phi^1 \rightarrow \phi^1 + 2 i p^0 \, m \;.
\end{eqnarray}

Now let us turn to the evaluation of  the integral
 \eqref{eq:Z-phi-theta}.
We begin with the following observation.
The $\theta$-dependent part of the exponential in \eqref{eq:Z-phi-theta}
reads ${\rm exp} [ i \pi \, \theta \, \varrho_a C^{ab} \varrho_b/ (p^0 p^1)] $, with $\varrho_a$ given in 
\eqref{rho-charges} \cite{Cardoso:2013ysa}.  
We will assume $|\varrho_a C^{ab} \varrho_b| \neq 0$, as the $\varrho_a C^{ab} \varrho_b= 0$ contributions
will be subleading.
Below we will evaluate the integral \eqref{eq:Z-phi-theta} by residue techniques.  To this end, we will 
first extend the range of integration to the entire real line and then 
move the contour to a region where the integrand becomes vanishing.
Choosing the contour specified by \eqref{attractorcontour}, we
obtain a non-vanishing result provided that $\varrho_a C^{ab} \varrho_b >0$.
The result will thus be proportional to 
a Heaviside step function $H(\varrho_a C^{ab} \varrho_b)$, as expected for an inverse Laplace
transform. Note that  $\varrho_a C^{ab} \varrho_b >0$ is a necessary condition for a 
charge configuration to correspond to a single-center black hole, as discussed below \eqref{rho-charges}.
Thus, in the following, we will use the contour \eqref{attractorcontour}.  It corresponds to the
so-called attractor contour introduced in  \cite{Cheng:2007ch} to single out large charge single-center contributions
to the entropy.

We now evaluate \eqref{eq:Z-phi-theta}. Here we proceed differently from 
\cite{Cardoso:2013ysa} and
resort to an approximation. Namely, 
we approximate the exact result by
only keeping the contributions  from
zeroes of $\Phi_{10}(\sigma (\theta), \rho (\theta), \theta)$
 that encode the dilatonic free energy contribution  \eqref{FD0}, as follows.
The zeroes of $\Phi_{10}(\sigma, \rho, v)$ that yield the leading contribution to the
entropy of single-center black holes are 
parametrized by three integers $(m, n, p)$ and given by \cite{Dijkgraaf:1996it},
\begin{equation}
\rho \sigma - v^2 + (1-2p) v + m \sigma - n \rho + p - p^2 - mn =0 \;.
\label{zeros-phi}
\end{equation}
This can be verified by considering a certain constrained extremization problem \cite{David:2006yn}.
In the following, we consider a related extremization problem in order to determine the subset of zeroes 
\eqref{zeros-phi} that encode the dilatonic free energy of single-center black holes.

To leading order, \eqref{eq:Z-phi-theta} can be calculated by saddle point approximation, by
extremizing the exponent on the right hand side of \eqref{eq:Z-phi-theta} with respect to $\theta$
subject to \eqref{zeros-phi}.
By inserting \eqref{eq:srvthet} into \eqref{zeros-phi}, we obtain the combination
\begin{equation}
D \equiv \rho (\theta) \sigma (\theta) - \theta^2 + (1-2p) \theta + m \sigma(\theta) - n \rho(\theta) + p - p^2 - mn
 = 0  \;, 
\label{Azeros}
\end{equation}
Denoting the exponent on the right hand side of \eqref{eq:Z-phi-theta} by $E$, 
\begin{equation}
E \equiv  - i \pi \sigma(\theta)\, q_a C^{ab} q_b + \pi q_a {\tilde \phi}^a (\theta) - \pi i \, \rho (\theta) \, p^a C_{ab} p^b \;,
\end{equation}
we 
consider the constrained extremization
problem,
\begin{equation}
\frac{d E}{d \theta} = \lambda \, \frac{d D}{d \theta} \;,
\label{AEl}
\end{equation}
where $\lambda$ denotes a Lagrange multiplier. 
We obtain
\begin{eqnarray}
\frac{d E}{d \theta} &=& i \pi \,  \frac{\varrho_a C^{ab} \varrho_b}{p^0 p^1} \;, \nonumber\\
\frac{d D}{d \theta} &=& \frac{i}{2} \left( \frac{\hat{\phi}^1}{p^1} - \frac{\hat{\phi}^0}{p^0} \right) - 2p
- m \frac{p^0}{p^1} + n \frac{p^1}{p^0} \;,
\label{EArho}
\end{eqnarray}
with ${\varrho}_a$ given in \eqref{rho-charges}.
Inserting this into \eqref{AEl} we get
\begin{equation}
\lambda = \frac{ 2 \pi \, \varrho_a C^{ab} \varrho_b}{p^0 \phi^1 - p^1 \phi^0} 
\end{equation}
as well as
\begin{equation}
\frac{p^1 n + l^0}{p^0} - \frac{(p^0 m  + l^1)}{p^1} = 2p \;.
\label{cond-mnp}
\end{equation}
The first relation determines the value of $\lambda$, while the second relation selects a subset of the zeroes \eqref{zeros-phi}.
The value of $\theta$ associated to these zeroes is determined from the condition $D =0$ given in 
\eqref{Azeros}, which we need to supplement with \eqref{cond-mnp}, resulting in 
\begin{eqnarray}
\label{value-theta-mnp}
2 \theta - 1 = - i \frac{\left(\phi^0 + 2i (l^0 + p^1 n)\right) \left(\phi^1 + 2i (l^1 + p^0 m)\right) + p^0 p^1  - 4 p^2 p^0 p^1}{
\phi^0 p^1 - \phi^1 p^0 } 
\;.
\end{eqnarray}
The exponent $E$, on the other hand, takes the value
\begin{equation}
E = \pi \left[ \frac{\hat{\phi}^0}{2 p^1} \, q_a C^{ab} q_b - \frac{\hat{\phi}^1}{2 p^0}\,  p^a C_{ab} p^b + q_a \phi^a
+ i \left(\theta - \frac12 \right) \frac{\varrho_a C^{ab} \varrho_b}{p^0 p^1} \right] \;,
\label{valueE}
\end{equation}
with $\theta$ given by \eqref{value-theta-mnp}. 

The zeroes \eqref{cond-mnp} depend on the combinations $l^0 + p^1 n$ and $l^1 + p^0 m$.
Using $\Phi_{10} (\sigma - n, \rho, v) = \Phi_{10} (\sigma, \rho + m, v) = \Phi_{10} (\sigma, \rho,v)$
as well as $q_a C^{ab} q_b \in 2 \mathbb{Z}, \, p^a C_{ab} p^b \in 2 \mathbb{Z}$, 
we can absorb the shifts $p^0 m$ and $p^1 n$ into $l^1$ and $l^0$ and extend the original range of $l^0$ and $l^1$ in \eqref{eq:Z-phi-theta}
to run over all the integers. 
The condition \eqref{cond-mnp} then becomes
\begin{equation}
\frac{l^0}{p^0} - \frac{l^1}{p^1} = 2p \;,
\label{cond-mnp-mod}
\end{equation}
where now $- \infty < l^{0,1} < \infty$.  Next, we parametrize the zeroes satisfying \eqref{cond-mnp-mod} by
\begin{equation}
(l^0, l^1, p) = ( (k + p) p^0,  (k - p) p^1, p) \;.
\label{zero-l-lp}
\end{equation}
The associated value of $\theta$ reads,
\begin{eqnarray}
2 \theta - 1 = - 2p + 2 k \frac{(\phi^0 p^1 + \phi^1 p^0 )}{\phi^0 p^1 - \phi^1 p^0 }
- i \frac{\phi^0 \phi^1   + p^0 p^1  - 4 k^2 p^0 p^1}{
\phi^0 p^1 - \phi^1 p^0 } 
\;.
\label{value-k-p}
\end{eqnarray}
Then, inserting \eqref{value-k-p} into $E$ in \eqref{valueE} shows that the real part of $E$ will depend on $k$, unless $k=0$, in which
case we obtain
\begin{equation}
e^E = e^{\pi [{\cal F}_D^{(0)} + q_a \phi^a]} \;,
\end{equation}
which is real and independent of $p$.
Here 
 ${\cal F}_D^{(0)}$ denotes the dilatonic
free energy introduced in \eqref{FD0}.  The zeroes with $k \neq 0$, on the other hand, 
correspond to instanton corrections to ${\cal F}_D^{(0)}$, 
\begin{equation}
\sum_k e^E = e^{\pi [{\cal F}_D^{(0)} + q_a \phi^a]} \, \sum_k e^{2 \pi i \tau \,  k^2 + 2 \pi i k z} \;,
\end{equation}
where in this expression $z \in \mathbb{R}$, whose value can be read off from \eqref{value-k-p}, 
and where $\tau$ is given by
\begin{eqnarray}
\tau = 2 i \frac{ \varrho_a C^{ab} \varrho_b}{(S + \bar S) |\phi^0 + i p^0|^2} \;.
\end{eqnarray}
Taking $S + \bar S >0$ as well as 
$\varrho_a C^{ab} \varrho_b >0$ (recall that this is implemented by using the contour (\ref{attractorcontour})), 
$\tau$ takes its value in the complex upper half plane, and the
sum over $k$ gives a theta function.

Thus, we have established that only the subset of zeroes parametrized by \eqref{zero-l-lp} solves
the extremization problem \eqref{AEl}, and that out of these only those with $k=0$ 
encode the dilatonic free energy ${\cal F}_D^{(0)}$, while those with $k \neq 0$ yield
instanton corrections.
In the following, and contrary to \cite{Cardoso:2013ysa}, we will suppress instanton corrections and only retain the subset of zeroes with $k=0$.  These are the zeroes
$(l^0, l^1, p) = (p p^0,  - p p^1, p)$. Since they contribute with a factor ${\cal F}_D^{(0)}$,
they yield the semi-classical free energy of a single-center $\frac14$ BPS black hole 
when $\varrho_a C^{ab} \varrho_b >0$.

We proceed to evaluate  \eqref{eq:Z-phi-theta} by retaining only the subset of zeroes just discussed,
and employing the contour (\ref{attractorcontour}).
Since $e^E$ is independent of $p$, and using the property $\Phi_{10} (\sigma, \rho, v +p)=
\Phi_{10} (\sigma, \rho, v)$ for $p \in \mathbb{Z}$, 
we can use
the zeroes $(l^0, l^1, p) = (p p^0,  - p p^1, p)$ to  extend the range of integration of $\theta_1$
to $- \infty < \theta_1 < \infty$. The relevant zero of $\Phi_{10}$ is then given by
${\cal D} = v + \rho \sigma - v^2 =0$.  In the vicinity of this zero, $\Phi_{10}$ takes the form
$\Phi_{10} \approx {\cal D}^2 \, \Delta$ with \cite{Jatkar:2005bh}
\begin{equation}
\Delta = \sigma^{-12} \, \eta^{24} (\gamma') \, \eta^{24}(\sigma') \;,
\end{equation}
where
\begin{equation}
\gamma' = \frac{\rho \sigma - v^2}{\sigma} \;\;\;,\;\;\; \sigma' = \frac{\rho \sigma - (v-1)^2}{\sigma} \;.
\end{equation}
In these expressions, 
$(\sigma, \rho, v)$ is replaced by \eqref{eq:srvthet} with $l^0 = l^1 =0$.
Then, using \eqref{valueE}, 
  \eqref{eq:Z-phi-theta} becomes
\begin{eqnarray}
\label{int-rep-01}
\sum_{q_0, q_1} d(q,p) \, {\rm e}^{\pi q_I \phi^I   } 
&=& \frac{1}{|p^0 p^1|}  
\,  \int_{-\infty}^{\infty} \dd \theta_1 
\, \frac{1}{{\cal D}^2(\theta) \, \Delta (\theta) } \\
&&  {\rm exp} \left[ 
 \pi \frac{\hat{\phi}^0}{2 p^1} \, q_a C^{ab} q_b - \pi \frac{\hat{\phi}^1}{2 p^0}\,  p^a C_{ab} p^b + q_a \phi^a
+ \pi i \left(\theta - \frac12 \right) \frac{\varrho_a C^{ab} \varrho_b}{p^0 p^1}
\right] \,. \nonumber
\end{eqnarray}
The contour of integration is
at fixed $\theta_2$ satisfying (\ref{attractorcontour}).
The quantity ${\cal D}^2 (\theta)$ has a double zero
at $2 \theta_* = 1 +  (S- \bar S)/(S + \bar S)$ \cite{LopesCardoso:2006bg}.
Recall that 
we consider configurations with  $\varrho_a C^{ab} \varrho_b \neq 0$.
We now evaluate the integral by residue technique, moving the contour to a region where
the integrand becomes vanishing.
In this way we find that only configurations with $\varrho_a C^{ab} \varrho_b > 0$ contribute.  They 
pick up the contribution from the zero ${\cal D} (\theta_*)=0$, resulting in 
\begin{eqnarray}
\sum_{q_0, q_1} d(q,p) \, {\rm e}^{\pi q_I \phi^I   }
=  \frac{M}{(S + \bar S)^2 |Y^0|^4} \, e^{\pi [{\cal F}_D + q_a \phi^a]} \;,
\label{sum01-free}
\end{eqnarray}
where
${\cal F}_D$ is the semi-classical dilatonic free energy in the presence of $R^2$ interactions,
\begin{eqnarray}
{\cal F}_D &=& {\cal F}_D^{(0)} + 4 \Omega (S, \bar S) \;, \nonumber\\
4 \pi \Omega(S, \bar S )&=& - \ln \eta^{24}(S) - \ln \eta^{24} (\bar S) - 12 \ln (S + \bar S) \;,
\label{eq:F-D}
\end{eqnarray}
while $M$ denotes the measure factor 
\begin{eqnarray}
\label{measure-M}
M &=&  H( \varrho_a C^{ab} \varrho_b ) \,  \Big[
\varrho_a C^{ab} \varrho_b  \\
&&- \frac{(S + \bar S)}{\pi} \Big(12 (Y^0 - \bar Y^0)^2 + (\ln \eta^{24}(S))' (S + \bar S) (\bar Y^0)^2
+ (\ln \eta^{24} (\bar S))' (S + \bar S)  (Y^0)^2 \Big)  \Big]\;.\nonumber
\end{eqnarray}
Here, $H$ denotes the Heaviside step function.  It ensures  that only configurations
with $\varrho_a C^{ab} \varrho_b > 0$ contribute.  As mentioned before, the latter is a
necessary condition for the charge configuration to constitute a single-center black hole.
We note that the 
Heaviside step function can be smoothen out into a continuous and differentiable function.
Below we will assume that this is the case, but will refrain from writing this out explicitly.

In obtaining these results we used that on the zero ${\cal D}(\theta_*) =0$ 
\cite{Cardoso:2013ysa},
\begin{eqnarray}
\sigma (\theta_*) &=& \frac{i}{S + \bar S} 
\;, \nonumber\\
\rho (\theta_*) &=& i \frac{|S|^2}{S + \bar S} \;.
\end{eqnarray}
Note that \eqref{sum01-free} no longer exhibits the shift symmetry $\phi^0 \rightarrow \phi^0 + 2i,
\phi^1 \rightarrow \phi^1 + 2i$, due to the fact that we only retained the contributions from
zeroes of $\Phi_{10}$ that give rise to the semi-classical dilatonic free energy ${\cal F}_D$.

Observe that both 
${\cal F}_D$ and the measure factor $M/[(S + \bar S)|Y^0|^2]^2$ in \eqref{sum01-free}
are invariant under S-duality transformations \eqref{Sduality}, \eqref{eq:electro-magn-dual_charges}.
This can be easily seen by rewriting $M$ in \eqref{measure-M} as 
\begin{eqnarray}
\label{M-Sd}
M = 2 \, H\left((T + \bar T)^a C_{ab} (T + \bar T)^b\right) 
 (S + \bar S) |Y^0|^2 \Big[ K + (S + \bar S)^2 \partial_S \partial_{\bar S} (4 \Omega) \Big] \;,
\label{measure-M-K}
\end{eqnarray}
with $T^a$ defined as in \eqref{Y0T}, and $K$ given in \eqref{K-Om}. The factor $M/[(S + \bar S)|Y^0|^2]$
is also T-duality invariant \cite{LopesCardoso:2006bg}.  The measure \eqref{measure-M-K}
is closely related to
(but not identical with) the measure factor $\sqrt{\Delta^-}$ introduced in 
\cite{LopesCardoso:2006bg} on
the grounds of electric/magnetic duality covariance.  It differs from $\sqrt{\Delta^-}$ 
by duality covariant terms.

Next, we would like to sum \eqref{sum01-free} over charges $q_a$ ($a=2, \dots, n$).  Here we face various issues.  First, we have a measure factor $M$ that depends on $q_a$.  To deal with this,
we first extend  $\phi^a$ to $\phi^a + i \mu^a$ (with $\mu^a \in \mathbb{R}^{n-1}$).  Then, we replace
the charge $q_a$ in $M$  by the corresponding differential operator of \eqref{diff-qp}.   This results in a differential
operator $\hat{M}$, which is obtained from $M$ by replacing $T^a$ with 
\begin{equation}
\hat{T}^a =  \frac{ \pi^{-1} \, C^{ab} \partial/\partial \mu^b +  {\bar S} p^a }{(S + \bar S) Y^0} \;.
\end{equation}
Thus, we replace \eqref{sum01-free} by
\begin{eqnarray}
\sum_{q_0, q_1} d(q,p) \, {\rm e}^{\pi [q_0 \phi^0 + q_1 \phi^1 + q_a (\phi^a + i \mu^a )]    }
= \frac{\hat{M}}{(S + \bar S)^2 |Y^0|^4} \, e^{\pi [{\cal F}_D + q_a (\phi^a + i \mu^a ) ]} \;,
\label{sum01-free-mu}
\end{eqnarray}
Next, we consider summing \eqref{sum01-free-mu} over $q_a$.
Here we face the problem that this sum
is ill-defined due to the indefinite signature of the $q_a$ charge lattice ($a=2, \dots, n$). 
Thus, the sum over $q_a$ has
to be regularized.  We propose the following procedure.  First, we pick a reference
vector $q_a^B$ such that $\varrho_a^B C^{ab} \varrho_b^B >0$.  
As mentioned below \eqref{Y-attrac}, we 
will refer to this
reference vector as an attractor background associated with a single-center BPS black hole.
We then consider fluctuations $V_a = q_a - q_a^B$ around this black hole background.
Thus, we set $q_a = q_a^B + V_a$ in \eqref{sum01-free-mu} 
and sum over $V_a$.  To enforce thermodynamic stability, we restrict to fluctuations
that do not increase the dilatonic free energy ${\cal F}_D$.  We do this by modifying the measure
factor in \eqref{sum01-free-mu}.  Namely, 
we introduce an additional
measure factor $\rho$, whose role is to weight each summand in the $V_a$ sum with $\pm 1$ or $0$, in such a way
that the contributions \eqref{sum01-free-mu} with growing exponent 
are removed from the sum, while 
the remaining contributions are weighted by $\pm 1$.  This is achieved by taking $\rho$ to be the difference of two sign functions
$\rho = \rho^{c_1} - \rho^{c_2}$, with $\rho^{c}(V; \tau) = - {\rm sgn} (V_a C^{ab} c_b) $ and suitably chosen vectors
$c_1$ and $c_2$ \cite{Zwegers:2008}.  Thus, the proposed regulator turns the sum over $V_a$
into an indefinite theta function based on sign functions.  In principle, we can also consider
indefinite theta functions that are 
based on error functions, as in \cite{Cardoso:2013ysa}.
These would then be defined in terms of different choices for $c_1$ and $c_2$.  Note that the regulator $\rho$
does not preserve all of T-duality, but only the subgroup $SO(1, n-2; \mathbb{Z})$.
We also note that there exist other proposals for regularizing the sum, which are based
on a modification of 
the exponent of \eqref{sum01-free-mu}.  Examples thereof are the so-called 
H-regulator, which has been proposed when $p^0=0$, and Siegel-Narain theta functions 
\cite{Dabholkar:2005dt,Gaiotto:2006wm,deBoer:2006vg,Denef:2007vg}. 

The resulting regularized partition function 
${\cal Z}_{\rm OSV}^{\rm reg} (p, \phi; \mu)$ appears to depend on 
the choice of the reference attractor background $q_a^B$, but this
dependence is only apparent, since two different 
choices of a reference background are related by an elliptic transformation of the indefinite
theta function.  Thus,  the result for the regularized partition function is independent of the choice of the background.  We proceed with the details of this construction.

The reference charge vector $q_a^B$ has a background value $\phi^a_B$  associated to it,
which is 
determined by \eqref{eq:phia-attrac}.
Expanding the exponent of \eqref{sum01-free-mu}
around $q_a = q_a^B + V_a$ 
gives 
\begin{eqnarray}
\pi \left[ {\cal F}_D(q) + q_a (\phi^a + i \mu^a) \right]
& = & 2\pi i \left[ \sfQ(q^B) \tau_e + \sfB(z_e, q^B) + \sfQ(V) \tau_e + \sfB(\Delta, V) \right]  \nonumber\\
&& + \pi \frac{ |S^2|}{S + \bar S} \, p^a \, C_{ab} \, p^b +  4\pi \Omega(S, \bar S)\;,
\end{eqnarray}
where we introduced
\begin{eqnarray}
\sfQ_e(q) &=& \frac12 q_a A^{ab} q_b \;\;\;,\;\;\; \sfB(z_e,q) = z_a^e A^{ab}  q_b \;\;\;,\;\;\; A^{ab} = - C^{ab} \;, 
\nonumber\\
\tau_e &=& \frac{i}{S+ \bar S} \;\;\;,\;\;\; 
z_a^e = \frac{i}{2} C_{ab} \left(\phi^b + i \frac{(S - \bar S)}{S + \bar S} p^b + i \mu^b \right) \;, 
\nonumber\\
\Delta_a &=& z_a^e + \frac{i}{S + \bar S} \,q_a^B = \frac{i}{2} C_{ab} \left(U^b 
+ i \mu^b \right) \;\;\;,\;\;\;U^a = \phi^a - \phi^a_B \;.
\end{eqnarray}
Then, by multiplying \eqref{sum01-free-mu} with the regular $\rho$ specified below and summing
over fluctuations $V_a \, (a = 2, \dots, n)$, we define the regularized partition function 
${\cal Z}_{\rm OSV}^{\rm reg} (p, \phi; \mu)$ by
\begin{eqnarray}
{\cal Z}_{\rm OSV}^{\rm reg} (p, \phi; \mu) \equiv
\frac{e^{\pi \frac{|S|^2}{S + \bar S} p^a C_{ab} p^b + 2  \pi i \sfQ (q^B) \tau_e + 4 \pi \Omega(S, \bar{S})
}}{(S+\bar{S})^2 |Y^0|^4}
 \, \hat{M} \,  \Big[ e^{ 2 \pi i \sfB (z_e, q^B) } \, 
 \vartheta(\Delta; \tau_e)  \Big] 
 \;,
\label{em-pf-reg2}
\end{eqnarray}
where $\vartheta(\Delta;\tau_e)$ denotes an indefinite theta function \cite{Zwegers:2008},
\begin{eqnarray}
\vartheta(\Delta;\tau_e) = \sum_{V\in \mathbb{Z}^{n-1}} \rho(V + \alpha; \tau_e) \, e^{2 \pi i \tau_e \sfQ_e(V) + 2\pi i \sfB(\Delta,V)} \;.
\label{ind-theta-Delta}
\end{eqnarray}
Here we decomposed 
$\Delta = \alpha \, \tau_e + \beta$ (with $\alpha, \beta \in \mathbb{Z}^{n-1}$), so that
\begin{equation}
\alpha_a = \frac12 (S + \bar S) C_{ab} U^b \;\;\;,\;\;\; \beta_a = - \frac12 C_{ab} \, \mu^b \;.
\end{equation}
The regulator $\rho = \rho^{c_1} - \rho^{c_2}$ is taken to be
\begin{equation}
\rho(q; \tau_e) = {\rm sgn}(\sfB(q,c_1)) - {\rm sgn}(\sfB(q,c_2)) \;,
\label{rho-zag-go}
\end{equation}
where ${\rm sgn}$ denotes the sign-function, and where $c_1$ and $c_2$ are two linearly independent null vectors,
$\sfQ(c_i) =0$. 

To assess the physical meaning of this construction, 
consider the case when $\phi^a$ does not equal $\phi_B$, but has a nearby value, so that
$U^a$ is small. 
A single unit of elliptic transformation shifts $\alpha_a$  by unity. Hence any given excitation about the specified  black hole background can be regarded as  a fluctuation characterized by $\alpha_a$ with $|\alpha_a|<1$. This puts the sum over the indefinite charge  lattices squarely in the domain of the G\"ottsche-Zagier treatment in \cite{GottZag1996},
as each component $\alpha_a$ is 
the range $0<|\alpha_a|<1$. Then, restricting
to a two-dimensional lattice $\Gamma^{1,1}$ for simplicity,
i.e. taking $n=3$, the indefinite theta function based on 
\eqref{rho-zag-go} precisely does what was described above, namely, contributions that would lead to an increase of the exponential in \eqref{ind-theta-Delta} are removed from the sum in a Lorentz invariant manner  \cite{GottZag1996}.  We refer to appendices B and D of 
\cite{Cardoso:2013ysa}
for a brief review of this.

Next, using \eqref{eq:phia-attrac}, we note that
the regularized partition function \eqref{em-pf-reg2} may also be written as
\begin{eqnarray}
{\cal Z}_{\rm OSV}^{\rm reg} (p, \phi; \mu) =
\frac{e^{\pi {\cal F}_E (\phi_B^a, p^a, S, \bar S)}}{(S+\bar{S})^{14} |Y^0|^4} \,
 \hat{M} \, \Big[ e^{2 \pi i \sfB(q^B, \Delta) } \, 
 \vartheta(\Delta; \tau_e)  \Big] \;,
 \label{exp-attr-back}
\end{eqnarray}
where ${\cal F}_E (\phi^a_B, p^a, S, \bar S)$ denotes the free energy \eqref{electric-free}
in the presence of $R^2$ corrections,
\begin{eqnarray}
\label{FE-hat-phi}
{\cal F}_E (\phi_B^a, p^a, S, \bar S) &=& - 2i \left(F(Y^a, S)\vert_{Y^a = \frac12 ({\phi}_B^a + i p^a)} - \bar{F} (\bar Y^a, \bar S)\vert_{\bar Y^a = \frac12 (\phi_B^a - i p^a)} \right) \;,
\nonumber\\
F(Y^a, S) &=& F^{(0)}(Y^a, S)  - \frac{i}{2\pi} \ln \eta^{24} (S) \;.
\end{eqnarray}
Note that in an $\cN = 2$ model, $F(Y^a, S)$ has the interpretation of a topological string free energy at weak
topological string coupling.
The
exponent ${\cal F}_E (\phi^a_B, p^a, S, \bar S)$  describes the semi-classical
free energy 
of a BPS black hole with charges $q^B_a$, while $\vartheta(\Delta;\tau)$ encodes the 
regulated contributions from the fluctuations $V_a = q_a - q_a^B$.

 As already mentioned, the choice of a reference background vector $q_a^B$ satisfying
 $\varrho_a^B C^{ab} \varrho_b^B >0$ is arbitrary.  Two different choices
 are related by an elliptic transformation of  $\vartheta(\Delta; \tau_e)$, as follows.
 Under the elliptic transformation
 $\Delta \rightarrow \Delta + \lambda \, \tau_e$
with $\lambda \in \mathbb{Z}^{n-1}$,  $\vartheta(\Delta; \tau_e)$ transforms as 
\begin{equation}
\vartheta(\Delta + \lambda \, \tau_e ;\tau_e) = e^{- 2  \pi i \sfQ (\lambda) \tau_e - 2 \pi i \sfB (\Delta, \lambda)}
\, \vartheta(\Delta; \tau_e) \;.
\label{ellipt-tran}
\end{equation}
Choosing two different reference background vectors $q^{B_1}$ and $q^{B_2}$ (both satisfying $\varrho_a^B C^{ab} \varrho_b^B > 0$ at fixed magnetic charges), and denoting the associated
values of $\Delta$ by $\Delta^1$ and $\Delta^2$, respectively, we implement
the elliptic transformation $\Delta^1 = \Delta^2 + \lambda \, \tau_e$ 
on  $\vartheta(\Delta^1; \tau_e)$, with
$\lambda = q^{B_1} -  q^{B_2}$.  This results in expression \eqref{em-pf-reg2}, with
$(q^{B_1}, \phi_{B_1})$ replaced by $(q^{B_2}, \phi_{B_2})$.  Thus, \eqref{em-pf-reg2}
is independent of the chosen background.
Here we have assumed that the
vectors $c_i$ which define $\rho$ are independent of any background value.

Since  \eqref{em-pf-reg2} is background independent, we may remove any reference to the
background $q^B$ by redefining the sum.  Rewriting  \eqref{em-pf-reg2} into a sum over charges $q_a = q_a^B + V_a$, we obtain
\begin{eqnarray}
{\cal Z}_{\rm OSV}^{\rm reg} (p, \phi; \mu) =
\frac{e^{\pi \frac{|S|^2}{S + \bar S} p^a C_{ab} p^b + 4 \pi \Omega(S, \bar{S})}}{(S+\bar{S})^2 |Y^0|^4 }  \, \hat{M} \, 
 \vartheta(z_e; \tau_e)  
\;,
\label{el-pf-reg}
\end{eqnarray}
where $\vartheta(z_e;\tau_e)$ denotes the indefinite theta function 
\begin{eqnarray}
\vartheta(z_e;\tau_e) = \sum_{q\in \mathbb{Z}^{n-1}} \rho(q + a; \tau_e) \, e^{2 \pi i \tau_e \sfQ_e(q) + 2\pi i \sfB(z_e,q)} \;.
\label{ind-theta}
\end{eqnarray}
Here we decomposed
$z_e$ as $z_e = a \tau_e + b$ (with $a, b\in \mathbb{R}^{n-1}$), resulting in 
\begin{equation}
a_a = \frac12 (S + \bar S) \,  C_{ab} \, \left( \phi^b  + i \frac{(S - \bar S)}{S + \bar S} p^b \right) \;\;\;,\;\;\; b_a = - \frac12 C_{ab} \, \mu^b \;.
\end{equation}
Note that since the measure factor $\hat{M}$ projects onto configurations with 
$\varrho_a C^{ab} \varrho_b > 0$, \eqref{el-pf-reg}  has the interpretation of a sum over attractor backgrounds. 
The regulated sum \eqref{el-pf-reg} has $\phi^a$-shift symmetry.  Namely, under
shifts of $\phi^a$ by $\phi^a \rightarrow \phi^a + 2i$, we have $C^{ab} z_b^e \rightarrow 
C^{ab} z_b^e - 1$, which leaves $\vartheta (z_e; \tau_e) $ invariant.

Let us now comment on a subtlety.
In the discussion below \eqref{rho-zag-go}
we took $\phi^a$ not to equal an attractor value $\phi_B^a$.
When $\phi^a$ is taken to be on an attractor value \eqref{eq:phia-attrac}, 
the components $a_a$ are integer valued and can be brought to zero by an appropriate elliptic transformation,
as discussed above.  In this case, the indefinite theta function \eqref{ind-theta-Delta} would vanish, unless we keep $\mu^a \neq 0$, so that $\Delta_a$ is non-vanishing. This subtlety was already noted in \cite{Cardoso:2013ysa}, and is the reason why
in \eqref{el-pf-reg} we have refrained from setting $\mu^a =0$ after the evaluation of $ \hat{M} \, 
 \vartheta(z_e; \tau_e)$.  
Then, by applying the modular transformation $\tau \rightarrow - 1/\tau$
(to be discussed below) we obtain a representation of the indefinite theta function with $a_a$ replaced by $- b_a = \frac12 C_{ab} \mu^b$.  Choosing $\mu^b$ so that $0 < |b_a| < 1$, we again obtain a set-up that is similar to the
one described below \eqref{rho-zag-go}.

Next, let us apply the modular transformation $(\tau_e, z_e) \rightarrow (-1/\tau_e, z_e/\tau_e)$ to 
\eqref{el-pf-reg} and discuss its consequences.
Taking into account that
$A^{ab}$ is integer valued, we obtain \cite{Zwegers:2008}
\begin{eqnarray}
\vartheta(z_e/\tau_e; - 1/\tau_e) &=& \frac{1}{\sqrt{- \det A}} \, (- i \tau_e)^{(n-1)/2} \, e^{2 \pi i \sfQ_e(z_e)/\tau_e}
\, \vartheta(z_e; \tau_e) \nonumber\\
&=& \sum_{\tilde{q}\in \mathbb{Z}^{n-1}} \rho(\tilde{q} + \tilde{a}; -1/\tau_e) \, e^{-2 \pi i \sfQ_e(\tilde{q})/\tau_e + 2\pi i \sfB(z_e/\tau_e,\tilde{q})} \;,
\label{ind-theta-transf}
\end{eqnarray}
where
\begin{equation}
\tilde{a} = \frac{{\rm Im}(z_e/\tau_e)}{{\rm Im}(-1/\tau_e)} = - b \;.
\end{equation}
Hence we get
\begin{equation}
 \vartheta(z_e; \tau_e) = \frac{\sqrt{- \det A}}{(- i \tau_e)^{(n-1)/2}} \, \sum_{\tilde{q} \in \mathbb{Z}^{n-1}}
 \rho(\tilde{q} - b; -1/\tau_e) \, e^{-2 \pi i \sfQ_e(\tilde{q} - z_e)/\tau_e} \;,
\label{ind-theta-transf2}
\end{equation}
where
\begin{equation}
{\tilde q}_a - z_a^e = - \frac{i}{2} C_{ab} \left(\hat{\phi}^b + i \frac{(S - \bar S)}{S + \bar S} p^b
\right) \;\;\;,\;\;\; \hat{\phi}^a = \phi^a + 2 i C^{ab} \tilde{q}_b + i \mu^a \;.
\end{equation}
Using \eqref{ind-theta-transf2}, we express \eqref{el-pf-reg} as
\begin{eqnarray}
\label{z-osv-el-reg}
&&{\cal Z}_{\rm OSV}^{\rm reg} (p, \phi; \mu) = 2 \sqrt{- \det A} \, (S + \bar S)^{(n-27)/2 } \, |Y^0|^{-2}
\\ 
 && H\left((\hat{T} + \bar{\hat{T}})^a C_{ab} (\hat{T} + \bar{\hat{T}})^b\right) 
 \left[ \hat{K} + 4 (S + \bar S)^2 \partial_S \partial_{\bar S} \Omega \right]
\, \sum_{\tilde{q} \in \mathbb{Z}^{n-1}}
 \rho(\tilde{q} - b; -1/\tau_e) \, e^{\pi \, {\cal F}_E ({\hat \phi}^a, p^a, S, \bar S)}
\;,
\nonumber
\end{eqnarray}
where ${\cal F}_E ({\hat \phi^a}, p^a, S, \bar S)$ denotes the free energy 
\eqref{FE-hat-phi}, with $\phi_B^a$ replaced by $\hat{\phi}^a$.

The OSV conjecture \cite{Ooguri:2004zv} relates the OSV partition function \eqref{OSV} to the topological free energy $F(Y^a,S)$ evaluated
at $Y^a = \frac12 (\phi^a + i p^a)$.  We proceed to extract this factor out of \eqref{z-osv-el-reg}, and obtain
\begin{eqnarray}
\label{z-osv-el-reg2}
&&{\cal Z}_{\rm OSV}^{\rm reg} (p, \phi; \mu) = 2 \sqrt{- \det A} \, (S + \bar S)^{(n-27)/2 } \, |Y^0|^{-2}
\\
&&
H\left((\hat{T} + \bar{\hat{T}})^a C_{ab} (\hat{T} + \bar{\hat{T}})^b\right) \,   \left[ \hat{K} + 4 (S + \bar S)^2 \partial_S \partial_{\bar S} \Omega \right]
\,  \, e^{\pi \, {\cal F}_E (\phi^a + i \mu^a, p^a, S, \bar S)}
\, \vartheta(z_e/\tau_e; - 1/\tau_e) \;. \nonumber
\end{eqnarray}
Thus, by making use of a modular transformation we have related
the regulated
sum ${\cal Z}^{\rm reg}_{\rm OSV}$ to the free energy ${\cal F}_E$.
The regularized partition function \eqref{z-osv-el-reg2} takes the form of an exponential factor
$|e^{- 2 \pi i F}|^2$, where $F$ denotes the holomorphic topological free energy \eqref{FE-hat-phi},
times a measure factor and an indefinite theta function.  If we artificially set $n=27$, which corresponds to taking a model with 28 abelian
gauge fields just as in the original ${\cal N}=4$ model, the powers of $S + \bar S$ cancel out in the measure
factor \cite{deWit:2007maa}, and we are left with 
the duality covariant differential operator 
$(\hat{K} + 4 (S + \bar S)^2 \partial_S \partial_{\bar S} \Omega)/|Y^0|^2$.  This operator, 
when acting on $e^{\pi {\cal F}_E}$,
yields the duality covariant factor $(K +  4 (S + \bar S)^2 \partial_S \partial_{\bar S} \Omega)/|Y^0|^2$, with
$K$ given in \eqref{K-Om} and $T^a = - i ({\phi}^a + i \mu^a + i p^a)/(2 Y^0)$.

\subsection{Summing over charges $p^a$}

Next, we turn to the black hole partition function \eqref{Z-enlarg-ens}, obtained by
summing over charges $p^a$ as well.
We take \eqref{sum01-free} as our starting point
and consider summing over both $q_a$ and $p^a$. Here we face the problem that
the dilatonic free energy ${\cal F}_D^{(0)}$ contains a term proportional to $(S - \bar S) q_a p^a$ that couples
one type of charges to the other type.  To avoid this coupling, we consider the case $S = \bar S$ in the following.
This allows us to interpret the sum over $q_a$ as a sum over attractor values $\phi^a_*$ at fixed $\chi_{a*}$, and the sum over $p^a$ as sum over attractor values $\chi_{a*}$ at fixed $\phi^a_*$, using \eqref{eq:phia-attrac}
and \eqref{eq:chia-attrac}.

We proceed as in the case of the regularized partition function \eqref{el-pf-reg}.
We first extend $\phi^a$ and $\chi_a$ to $\phi^a + i \mu^a$ and $\chi_a + i \nu_a$, respectively (with $\mu^a, \nu_a \in \mathbb{R}^{n-1}$).  We convert the measure $M$ in \eqref{measure-M-K}
into a differential operator $\hat{M}$, obtained by replacing $T^a$ by the differential operator \eqref{hat-T}.
Then, taking \eqref{sum01-free} as a starting 
point, we obtain
\begin{equation}
\sum_{q_0, q_1} d(q,p) \, e^{\pi [q_0 \phi^0 + q_1 \phi^1 + q_a (\phi^a + i \mu^a) - p^a ( \chi_a
+ i \nu_a ) ] } = 
\frac{\hat{M}}{ 4 S^2 |Y^0|^4}  \, e^{\pi [{\cal F}_D + q_a (\phi^a  + i \mu^a) - p^a (\chi_a + i \nu_a) ]}
\;.
\label{sum01-free-mu-nu}
\end{equation}
Then, in analogy to  \eqref{el-pf-reg}, 
we define the regularized partition function, obtained by summing \eqref{sum01-free-mu-nu} 
over charges $q_a$ and $p^a$ ($a = 2, \dots, n$),
by
\begin{eqnarray}
{\cal Z}^{\rm reg} (\phi^I, \chi_a; \mu, \nu) \equiv
\frac{e^{4 \pi \Omega(S)}}{4 S^2 |Y^0|^4 } \,
\hat{M} \, 
\Big( \vartheta(z_e; \tau_e) \, \vartheta(z_m; \tau_m) \Big) 
\;,
\label{em-pf-reg}
\end{eqnarray}
where $\vartheta(z_e; \tau_e)$ and  $\vartheta(z_m; \tau_m)$ denote indefinite theta functions with 
\begin{eqnarray}
\tau_e &=& \frac{i}{2 \,S}  \;\;\;,\;\;\; \tau_m = \frac{i}{2} \,  S \nonumber\\
z_a^e &=& \frac{i}{2} C_{ab} \left(\phi^b + i \mu^b \right) \;\;\;,\;\;\; z^a_m = - \frac{i}{2} C^{ab} \left(\chi_b + i \nu_b \right) \;, \nonumber\\
\sfQ_e(q) &=& \frac12 q_a A^{ab} q_b \;\;\;,\;\;\; \sfQ_m(p) = \frac12 p^a A_{ab} p^b \;\;\;,\;\;\;
A_{ab} = - C_{ab} \;\;\;,\;\;\; A^{ab} = - C^{ab} \;, \nonumber\\
\sfB(z_e, q) &=& z_a^e A^{ab} q_b \;\;,\;\;\; \sfB(z_m, p) = z_m^a A_{ab} p^b \;.
\end{eqnarray}
We take both indefinite theta functions to be defined in terms of sign functions, as in 
\eqref{rho-zag-go}.  
Observe that \eqref{em-pf-reg} is invariant under $S \rightarrow 1/S$. It is also invariant under 
$SO(1,n-2;\mathbb{Z})$ T-duality transformations, as well as under shifts
$\phi^a \rightarrow \phi^a + 2i$ and 
$\chi_a \rightarrow \chi_a + 2i$.

Applying the modular transformations $(\tau_e, \tau_m) \rightarrow (-1/\tau_e, -1/\tau_m)$ we obtain, in a manner
analogous to \eqref{ind-theta-transf}, 
\begin{equation}
\vartheta(z_e; \tau_e) \, \vartheta(z_m; \tau_m) =  2^{n-1} \sum_{{\tilde q} \in \mathbb{Z}^{n-1}} 
\sum_{{\tilde p} \in \mathbb{Z}^{n-1}} \rho(\tilde{q} - b^e; - 1/\tau_e)  \, \rho(\tilde{p} - b_m; - 1/\tau_m) \,
e^{\pi {\cal H}^{(0)}(S, \hat{\phi}^a, \hat{\chi}_a)} \;,
\end{equation}
where 
\begin{eqnarray}
\hat{\phi}^a &=& \phi^a + 2 i C^{ab} \tilde{q}_b + i \mu^a \;\;\;,\;\;\;
\hat{\chi}_a = \chi_a - 2 i C_{ab} \tilde{p}^b + i \nu_a \;, \nonumber\\
b^e_a &=& - \frac12 C_{ab} \mu^b \;\;\;,\;\;\; b_m^a = \frac12 C^{ab} \nu^b \;,
\end{eqnarray}
and where $ {\cal H}^{(0)}$ denotes the Hesse potential \eqref{hesse-a}. Using this, we arrive at 
\begin{eqnarray}
{\cal Z}^{\rm reg} (\phi^I, \chi_a; \mu, \nu) = 2^{n-3}
\frac{\hat M}{S^2 |Y^0|^4 } \,
\Big( e^{\pi {\cal H} (S, \phi^a + i \mu^a, \chi_a + i \nu_a)} \, \vartheta(z_e/\tau_e; -1/\tau_e) \, 
\vartheta(z_m/\tau_m; -1/\tau_m) \Big) ,\nonumber\\
\end{eqnarray}
where we introduced the Hesse potential ${\cal H}$ in the presence of higher-derivative corrections 
\cite{LopesCardoso:2006bg},
\begin{equation}
{\cal H} (S, \phi^a, \chi_a) = {\cal H}^{(0)} (S, \phi^a, \chi_a) + 4 \Omega (S) \;.
\end{equation}
Thus, by resorting to modular transformations, we have related the regularized partition function 
\eqref{em-pf-reg} to the semi-classical Hesse potential.

Now let us expand \eqref{em-pf-reg} around an attractor background defined in terms of charges $(q_a^B, p^a_B)$ satisfying $\varrho_a^B C^{ab} \varrho_b^B > 0$. 
The associated background values $(\phi^a_B, \chi_a^B)$ are given by \eqref{eq:phia-attrac} and 
\eqref{eq:chia-attrac}, where we recall that we are setting $S = \bar S$.  Hence, the background value
$\phi^a_B$ is determined in terms of $q_a^B$ (and $S$), while $\chi_a^B$ is determined in terms of 
$p^a_B$ (and $S$).  The choice of the background values  $(q_a^B, p^a_B)$ is arbitrary.
We perform the shifts
\begin{eqnarray}
q_a &=& q_a^B + V_a \;\;\;,\;\;\; p^a = p^a_B + W^a \;, \nonumber\\
\phi^a &=& \phi_B^a + U^a \;\;\;,\;\;\; \chi_a = \chi_a^B + X_a  \;.
\label{backgr-exp-qp}
\end{eqnarray}
Expanding \eqref{em-pf-reg} around $(q_a^B, p^a_B)$ is implemented by applying the following elliptic transformations
to $\vartheta(z_e; \tau_e)$ and  $\vartheta(z_m; \tau_m)$,
\begin{equation}
z^e_a = {Z}^e_a - q_a^B \, \tau_e \;\;\;,\;\;\; z_m^a = {Z}_m^a - p_B^a \, \tau_m \;,
\end{equation}
where
\begin{equation}
{Z}^e_a =  \frac{i}{2} C_{ab} \left(U^b + i \mu^b \right) \;\;\;,\;\;\; 
{Z}^a_m = - \frac{i}{2} C^{ab} \left(X_b + i \nu_b \right) \;.
\end{equation}
Using the transformation property \eqref{ellipt-tran} gives
\begin{eqnarray}
{\cal Z}^{\rm reg} (\phi^I, \chi_a; \mu, \nu) &=&
\frac{e^{\pi {\cal H} (S, \phi_B, \chi^B) + \pi q_a^B U^a-
\pi p^a_B X_a}}{4 \, S^2 \, |Y^0|^4 } \,
\nonumber\\
&& \qquad  \, \hat{M} \, 
\Big( 
e^{\pi i q_a^ B \mu^a - \pi i p^a_B \nu_a  } \,
\vartheta({Z}_e; \tau_e) \,
\, \vartheta({Z}_m; \tau_m) 
\Big)  
\;,
\label{gcp-reg}
\end{eqnarray}
where
\begin{eqnarray}
\vartheta({Z}_e; \tau_e) &=& \sum_{V \in \mathbb{Z}^{n-1}} \rho( V + {A}_e; \tau_e)
\, e^{2 \pi i \tau_e  \sfQ(V ) + 2 \pi i   \sfB( {Z}_e, V )} \;, \nonumber\\
\vartheta({Z}_m; \tau_m) &=& \sum_{W \in \mathbb{Z}^{n-1}} \rho( W + {A}_m; \tau_m)
\, e^{2 \pi i \tau_m  \sfQ(W ) + 2 \pi i  \sfB( {Z}_m, W )} \;.
\end{eqnarray}
Here we decomposed ${Z} = {\cal A} \tau + {\cal B}$, so that 
\begin{eqnarray}
{\cal A}_a^e &=& S \, C_{ab} U^b  \;\;\;,\;\;\; {\cal B}_a^e = - \frac12 C_{ab} \mu^b \;,
\nonumber\\
{\cal A}^a_m &=&
- \frac{1}{S}  C^{ab} X_b  \;\;\;,\;\;\; {\cal B}_m^a =  \frac12 C^{ab} \nu_b \;.
\end{eqnarray}
In \eqref{gcp-reg}, the first line gives the contribution of the attractor background to the partition
function, while the second line contains the contribution from fluctuations around it.

Observe that the final expression \eqref{gcp-reg} is identical to what one obtains starting from 
the regularized OSV partition function \eqref{exp-attr-back}, with $S = \bar S$, multiplying it
with $e^{- \pi p^a (\chi_a + i \nu_a)}$, summing over charges $p^a$ by resorting to the background expansion
\eqref{backgr-exp-qp} and regularizing this sum.
Thus, our proposal \eqref{em-pf-reg} for the regularized partition function is consistent with
what one obtains by first regularizing the sum over $q_a$, which results in \eqref{exp-attr-back},
and subsequently
summing over the charges $p^a$ and regularizing this sum in a similar manner.
Note that our proposal \eqref{em-pf-reg} does not depend on any particular
attractor background $(\phi_B, \chi^B)$.

Observe that the regularized partition function \eqref{em-pf-reg} only counts axion-free attractor backgrounds
\eqref{eq:phia-attrac} and
\eqref{eq:chia-attrac}.  We may extend this by implementing the S-duality transformation $S \rightarrow S + i$ and
summing over all its images.

Finally, we note that
the form of \eqref{em-pf-reg} is reminiscent of the proposal \cite{Eynard:2008he} for a background independent
partition function for matrix models and topological strings.

\section{Conclusions}\label{discussion}

We first defined an OSV partition function for single-center BPS black holes by restricting
to those zeroes of $\Phi_{10}$ that give rise to the dilatonic free energy ${\cal F}_D$ of
single-center BPS black holes.  Within this approximation, we dealt with the divergences
that arise when performing the sum over charges $q_a$ by first expanding around an attractor background,
and then regularizing the sum over fluctuations $V_a$ around this background
 by removing contributions that
are exponentially growing.  This was achieved by converting the $V_a$-sum into
an indefinite theta function based on a regulator $\rho$ constructed out of sign functions.
The choice of the attractor background is arbitrary, and
two different choices are related by an elliptic transformation of the indefinite theta function.
We then used its modular
properties to relate the regularized sum to the free energy ${\cal F}_E$ which, in $\cN=2$
compactifications, is relatd to the topological free energy at weak topological
string coupling. 
Observe that the regulator $\rho$ 
only preserves a subset of T-duality transformations, namely the one given by $SO(1, n-2; \mathbb{Z})$ transformations.

We then turned to a canonical partition function by also summing over charges $p^a$.
We set $S = \bar S$ in order to decouple the sums over $q_a$ and $p^a$. We again
regularized the sums using indefinite theta functions. We then used a modular transformation
to relate the regularized canonical partition function to the Hesse potential $\cal H$ of
supergravity. The regularized sum is invariant under the electric-magnetic duality
transformation $S \rightarrow 1/S$. Its
form is reminiscent of the proposal \cite{Eynard:2008he} for a background independent
partition function for matrix models and topological strings.

One subtlety that arises in our proposal is that if we choose $\phi^a$ and/or $\chi_a$ to be
on an attractor value \eqref{eq:phia-attrac} and/or \eqref{eq:chia-attrac}, 
the indefinite theta functions vanish unless we extend $\phi^a, \chi_a$
to the complex plane, i.e. $\phi^a \rightarrow \phi^a + i \mu^a, \, \chi_a \rightarrow \chi_a + i \nu_a$.  Thus, our regularized partition functions depend on $\mu^a, \nu_a$.

We chose a regulator with good modular and elliptic transformation properties in order to
be able to relate the regularized partition function to semi-classical results, and to ensure that
the partition function is independent of the particular attractor background around which one
chooses to expand it.  We opted to work with indefinite theta functions based on sign functions,
but other choices are, in principle, also possible \cite{Dabholkar:2012nd,Cardoso:2013ysa}.

Finally, we note that the computation of a Witten index in the presence of a continuous
spectrum may yield a result \cite{Akhoury:1984pt,Giveon:2014hfa,Ashok:2014nua} that is reminiscent of an indefinite theta function.
Consider a supersymmetric one-dimensional quantum mechanics model with Hamiltonian 
$H = 
p^2 + W^2 (x) - [\psi^\dagger , \psi] \, W' (x) $.
This Hamiltonian describes a charged spin $\frac12$ particle moving in a potential $W (x)$.
Take $W$ to have a solitonic form, i.e. 
\begin{equation}
\left\{
\begin{matrix}
W (x) \longrightarrow W_+
& \qquad \text{for} \, x \longrightarrow + \infty \\
W (x) \longrightarrow W_-
& \qquad \text{for} \, x \longrightarrow - \infty
\end{matrix} \right. \ .
\end{equation}
Upon imposing boundary conditions, an explicit computation of the Witten index in this model yields 
\cite{Akhoury:1984pt}
\begin{equation}
E (\sqrt{\beta} \, W_+
) - E (\sqrt{\beta} \, W_-) \;,
\end{equation}
where $E$ denotes the error function.  In the presence of superselection sectors
labelled by $\gamma$, this generalizes to 
\begin{equation}
Z (\beta ) = \sum_{\gamma} \left( E (\sqrt{\beta} \, W_+^{\gamma}) - E (\sqrt{\beta} \, W_-^{\gamma}) \right) \ \e^{- \beta H_{top}^{\gamma}} \;,
\end{equation}
where we allowed for the presence of a topological term $H_{top}$ in the Hamiltonian \cite{Denef:2011ee}.
Then, a 
judicious choice of both the asymptotics of the potential $W_{\pm}^{\gamma}$ and the topological Hamiltonian $H_{top}^{\gamma}$ labeling the superselection sectors, gives an indefinite theta function.

\vskip 5mm

\subsection*{Acknowledgements}
We acknowledge valuable discussions with Kathrin Bringmann, Atish Dabholkar, Jan Manschot, Sameer Murthy, Daniel Persson, Larry Rolen, Jan Troost and Stefan Vandoren.
The work of G.L.C. and M.C. is
partially funded by Funda\c{c}\~{a}o para a
Ci\^{e}ncia e a Tecnologia (FCT/Portugal) through project
PEst-OE/EEI/LA0009/2013 and through grants PTDC/ MAT/119689/2010
and EXCL/MAT-GEO/0222/2012. 
M.C. is also supported by FCT/ Portugal via the Ci\^{e}ncia 2008 program. 
S.N. is supported by a URC fellowship, an NRF grant and
an SARCHI fellowship.
This work is also partially supported by the COST action
MP1210 {\it The String Theory Universe}.

\end{document}